# First Results of the Magnetometer (MAG) Payload onboard Aditya-L1 Spacecraft

Vipin K. Yadav*[1], Y. Vijaya[2], P. T. Srikar[3], B. Krishnam Prasad[2], Monika Mahajan[2], K. V. L. N. Mallikarjun[2], S. Narendra[3], Abhijit A. Adoni[3], Vijay S. Rai[3], D.R. Veeresha[3], and Syeeda N. Zamani[2]

(1) Space Physics Laboratory (SPL), Vikram Sarabhai Space Centre (VSSC), Thiruvananthapuram 695022, India
(2) Laboratory for Electro-Optical Systems (LEOS), ISRO, Peenya, Bengaluru 560058, India
(3) U R Rao Satellite Centre (URSC), Old Airport Road, Vimanpura, Bengaluru 560017, India

* Email: vipin_ky@vssc.gov.in

## Abstract

Aditya-L1 is the first Indian solar mission placed at the first Lagrangian (L1) point to study the Sun. A fluxgate magnetometer (MAG) is one of the seven payloads and one of the three *in-situ* payloads onboard to measure the interplanetary magnetic field (IMF) coming from the Sun towards the Earth. At present, the Aditya-L1 spacecraft is in a halo-orbit around the L1 point and the MAG payload is ON is continuously measuring the IMF. This paper presents the first measurements of the IMF by MAG.

## 1. Introduction

Magnetic field experiments are regularly sent onboard space missions so as to estimate the nature and magnitude of the interplanetary magnetic field (IMF) emanating from the Sun in various heliospheric regions. These IMF measurements are crucial in understanding the space weather phenomena in heliosphere at various places which include the Lagrangian points also.

**1.1 The Interplanetary Magnetic Field (IMF)**

The IMF is the solar magnetic field which is brought into the heliosphere by the solar wind plasma emanating from the Sun. From the solar photosphere, the IMF comes out from a region where the magnetic field lines are not closing back at a nearby region but instead travels deep in the heliosphere before returning back towards the Sun for closing. The black sunspots on the Sun's photosphere are the regions from where the IMF originates. The direction of IMF in the northern hemisphere of Sun is opposite to that of the IMF in the southern hemisphere of Sun. This magnetic field polarity reverses with the solar cycle.

**1.2 Importance of Magnetic Field Measurements at L1 Point**

The scientific importance of the IMF measurements at L1 point is as follows:

1) These measurements provide an insight about the IMF coming towards Earth from the Sun.
2) The IMF observation at L1 point helps in understanding the solar wind particle distributions reaching at the L1 point from the Sun.
3) These IMF measurements can provide inputs in identifying the solar wind source locations for the thermal and energetic particles which follow the IMF during propagating in heliosphere.
4) The IMF observations at L1 point support the detection of solar plasma wave signatures.

**1.3 Earlier Missions at and around L1 Point**

Six missions such as ISEE-3, WIND, SOHO, ACE, and GENESIS have been sent to L1 point or near the L1 point such as STEREO to study the Sun and the solar wind. Primarily, in these spacecrafts imaging instruments and the charged particle (electron, proton, and ion) detectors were installed. However, three missions namely ISEE-3, WIND and ACE, also carried magnetometers onboard for measuring the IMF.

## 2. The Aditya-L1 Mission

Aditya-L1 is the first Indian solar mission to continuously observe the sun from the first Lagrangian (L1) point [1]. The Aditya-L1 spacecraft is placed into a halo-orbit around the L1 point at a distance of $\approx 1.5 \times 10^6$ km away from the Earth towards the Sun on the Sun - Earth line. The Aditya-L1 spacecraft has seven payloads onboard to continuously observe the Sun. Four payloads onboard namely VELC, SUIT, SoLEXS and HEL1OS in different visible and ultraviolet (UV) wavebands are designed to study the Sun and the solar corona in visible, ultraviolet and x-rays [2]. ASPEX, PAPA and MAG are *in-situ* experiments to measure the solar wind particles [3].

**2.1 Science Objectives of the MAG Payload**

The MAG payload is to study the variability in IMF at L1 point so that the quiet Sun background magnetic field can be estimated. The IMF influences the acceleration of solar wind particles and the transport of energy in heliosphere. The IMF measurements by MAG shall complement the ASPEX and PAPA observations. The science objectives of MAG [4] are:

1. To measure the magnitude and the direction of the IMF reaching at L1 point from the Sun and its variation with time.
2. To identify and observe the passage of the Coronal Mass Ejections (CMEs) structure at L1 point.
3. To provide inputs for the space weather studies as to how the extreme solar events are affecting the earth's magnetosphere.

4. To detect the signatures of the solar plasma waves reaching the L1 point and coming towards the Earth.

## 3. The MAG Payload

The MAG is designed to measure the IMF at L1 point which typically varies between 5-10 nT in a solar day but can go up to 50 nT or even more during solar transient events.

The MAG payload consists of two sensors - MAG-11 and MAG-12 having 3-orthogonally mounted fluxgate sensors, and a processing electronics package MAG-13. The two fluxgate sensors are mounted on a 6 m deployable boom; one sensor is mounted at a distance of 3 m, while the other at 6 m at the tip of the boom so that both the sensors measure the same IMF but the different spacecraft magnetic field which can be removed by invoking a differential. The MAG fluxgate sensor is realized by winding an excitation coil on a Supermalloy ring core in a toroidal assembly and placing it in a MACOR bobbin on which a sensing coil is wound on top completely enclosing the excitation winding.

The processing MAG-13 is placed inside the spacecraft mainframe. The MAG electronics module consists of two analog processing boards (one for each sensor), a common digital processing board, EMI-filter, triple output DC-DC converter, and an interface board facilitating all the interconnections between various boards. The flight model (FM) of MAG payload is showed in Figure 1.

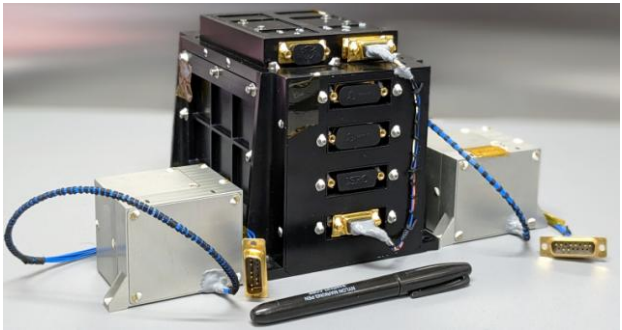

**Figure 1.** MAG payload flight model (FM).

The MAG sensors are designed to measure the IMF vectors in the range of ±256 nT per axis once in every 128 ms with a resolution of 0.1 nT and accuracy of 0.5 nT. MAG has other magnetic field ranges of ±5000 nT, ±10,000 nT and ±60,000 nT so that the payload can be tested on ground in the presence of geomagnetic field, the payload can be switched on during the cruise phase and a backup for the saturation of sensors in halo-orbit, etc [5].

The boom for MAG payload is designed so that the sensitive fluxgate sensors can be placed away from the spacecraft in order to prevent them from saturation due to the magnetic field generated by the Aditya-L1 as a whole. The MAG boom is realized with carbon fiber reinforced polymer (CFRP) and has four arm of 1.5 m long and another of 0.75 m length so that the boom can be deployed in the – roll (downward) direction. The boom can hold two sensors of 200 gm each and is built with a mechanism so that after the deployment the boom arms shall be locked and cannot be folded back. The payload electric signal and the harness wires are routed on the rectangular boom tubes employing coaxial cables. The MAG payload with boom installed onboard the Aditya-L1 spacecraft is shown in Figure 2.

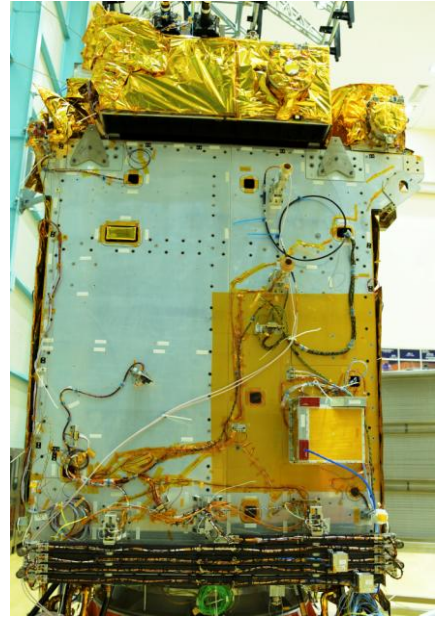

**Figure 2.** MAG payload onboard Aditya-L1 spacecraft.

The MAG payload onboard Aditya-L1 is configured to meet the science objectives of measuring the IMF. The technical specifications of MAG are given in Table 1.

| Parameter | Value |
|---|---|
| Range, Resolution, Accuracy (in nT) 4 ranges | ± 60,000   2.0   8.0<br>± 10,000   0.4   2.0<br>± 5,000    0.2   1.0<br>± 256      0.1   0.5 |
| Sampling rate | 8 vectors/s, 16 bit/sample |
| Spacecraft Interface | MIL-STD-1553B |
| Operating Temperature (in °C) | Sensor: - 35 to + 65<br>Electronics: - 15 to + 55 |
| Non-linearity | 0.7\% max. (of full scale in all ranges) |
| Total Mass (kg) | 15.6 [Sensors: 0.5, Electronics: 2.5; Boom: 11.1; Harness/Thermal: 1.5] |
| Raw Power | 14.64 W [7.64 + 7.0 (1553 transaction)] |
| Size (Sensor) | 87 mm × 66 mm × 56 mm |
| Size (Electronics) | 170 mm × 125 mm × 117 mm |
| Boom Length | 6 m |

The sources of error in the MAG payload include both the sensor and the electronics. All the possible error sources for MAG are identified and the total uncertainty in MAG measurements is estimated to be less than 0.4 nT.

## 4. First Observations by MAG in Space

Aditya-L1 was launched from Sriharikota on September 2, 2023 and the MAG payload was switched ON for the first time on October 16, 2023 when the boom was in the stowed condition and since then the MAG payload was ON during the cruise phase from Earth to L1 point in the magnetic field range of ± 10,000 nT though both MAG sensors were at high temperatures of 60$^o$C and 50$^o$C.

### 4.1 MAG Observations with stowed boom

During the cruise phase the magnetic contamination got stabilized so that it was subtracted from the total magnetic field as measured by the MAG sensors. On November 05, 2023 the MAG payload observed a CME on which shows a significant increase in the IMF. The IMF, captured by the MAG sensors, was compared with the observations made by the magnetometer onboard DSCOVR spacecraft and are shown in Figure 3 which shows that MAG picked all the prominent features in the IMF variation and overlapping with DSCOVR measurements.

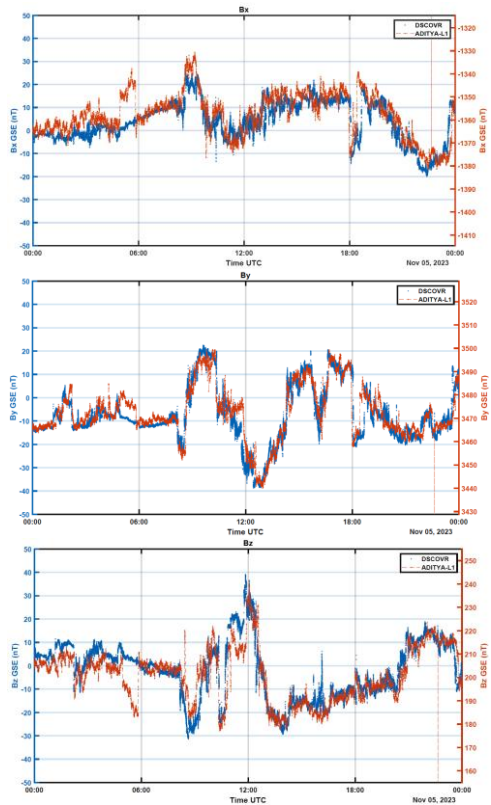

**Figure 3.** IMF observations by MAG onboard Aditya-L1 (red) en-route to L1 and DSCOVR spacecraft (blue).

### 4.2 MAG Observations after boom deployment

The Aditya-L1 spacecraft was inserted into a halo-orbit on January 06, 2024 and subsequently, the MAG boom is deployed on January 11, 2024. After the boom deployment, the temperatures of both the MAG sensors came down at a comfortable level viz. MAG-11 from 64$^o$C to 17$^o$C and MAG-12 from 55$^o$C to 16$^o$C.

After the boom deployment, the MAG sensor range was reduced to ±5,000 nT from ±10,000 nT and the magnetic field measurements were carried out for an Earth day. No saturation in any axis of any sensor was observed and hence, the MAG sensor range was further reduced to ±256 nT and since then both the MAG sensors are operating in this range with any saturation.

After the boom deployment, once again the magnetic field measurements by both MAG sensors were made independently and were compared with the DSCOVR observations which are shown in Figure 4 for MAG-11 sensor and in Figure 5 for MAG-12 sensor.

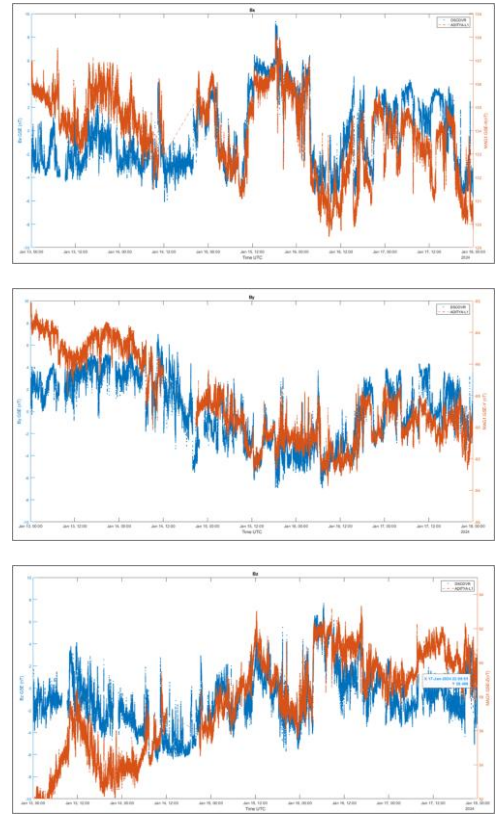

**Figure 4.** MAG-11 observations (red) after the boom deployment and in comparison with DSCOVR (blue).

It is to be noted that these are uncorrected observations where the axis offsets are not accounted for.

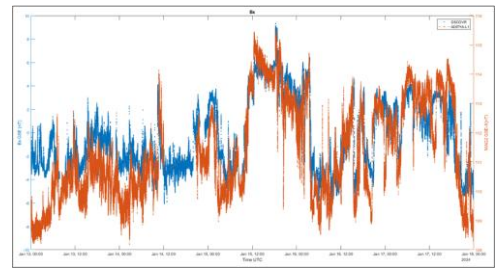

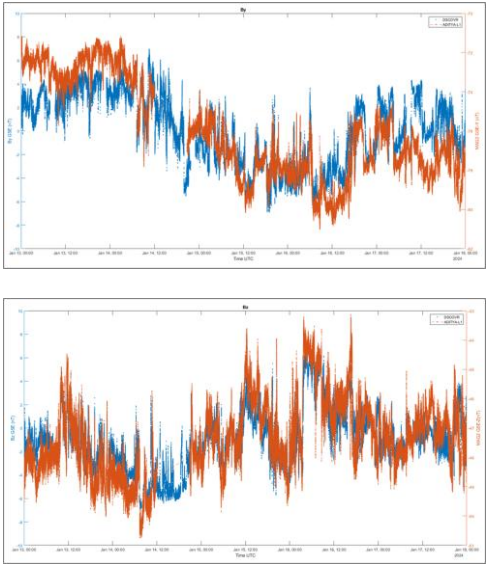

**Figure 5.** MAG-12 observations (red) after the boom deployment and in comparision with DSCOVR (blue).

After the boom deployment, measurements are regularly carried out and the plots for January 21, 2024 indicate a general trend in each field component of MAG-1 and MAG-2 in 256 nT as shown in Figure 6 which indicated that the sensor noise and the magnetic field from spacecraft is embedded within this and is not separable from IMF as long as the axis offsets are not estimated.

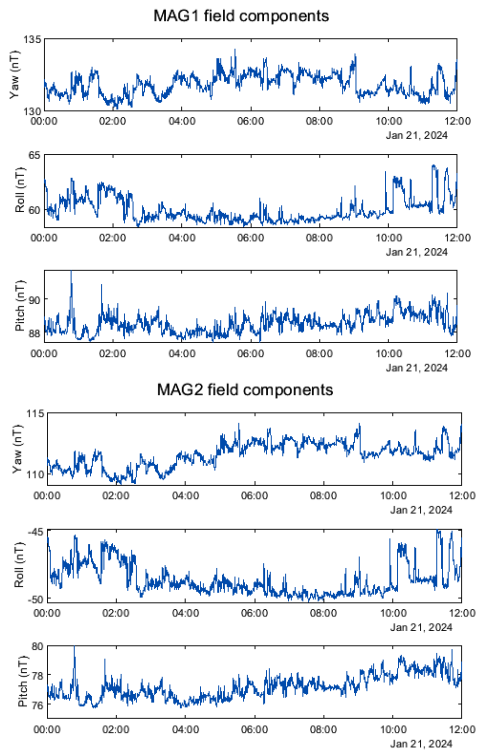

**Figure 6.** Sample plot with all the three components for MAG1 and MAG2 sensors observed on January 21, 2024.

The spacecraft rotation was performed on April 22, 2024 about the yaw axis so that the y- and z-axis offsets for both the sensors can be estimated. The spacecraft was rotated away off-axis from the Sun line on May 01, 2024 and the x-axis offset was estimated. With the axis offset estimates in hand, the MAG payload verification activities are presently going on. The IMF in the halo-orbit around the L1 point by MAG payload onboard Aditya-L1 spacecraft are being obtained after invoking the differential between the two sensor sets.

## 6. Concluding Remarks

The MAG payload onboard Aditya-L1, the first Indian solar mission is one of the seven payloads onboard. After its launch in September, 2023, the MAG payload sensors were powered in October, 2023 and since then the MAG payload is ON. The MAG measured the magnetic field en-route to L1 point with boom in the stowed condition. The Aditya-L1 spacecraft was inserted in the halo-orbit around the L1 point in January, 2024 and subsequently the boom was deployed and since then the MAG sensors are continuously measuring the IMF and presently the payload verification (PV) phase activities are going on. During this various sets of IMF are captured by the MAG payload which indicated it's satisfactory and the scientific performance on the expected lines. In this duration, few solar extreme events are also picked by the MAG sensors.